\documentclass[a4paper,11pt]{article}
\usepackage{pos}

\renewcommand{\logo}{\relax}

\usepackage{subcaption}

\title{Persistent homology as a probe for center vortices and deconfinement in SU(2) lattice gauge theory}
\ShortTitle{Persistent homology as a probe for center vortices in SU(2) lattice gauge theory}

\author*[a]{Nicholas Sale}
\author[a,b]{Biagio Lucini}
\author[c]{Jeffrey Giansiracusa}

\affiliation[a]{Department of Mathematics, Swansea University, Bay Campus, SA1 8EN, Swansea, Wales, UK}
\affiliation[b]{Swansea Academy of Advanced Computing, Swansea University, Bay Campus, SA1 8EN, Swansea, Wales, UK}
\affiliation[c]{Department of Mathematical Sciences, Durham University, Upper Mountjoy Campus, Durham, DH1 3LE, UK}

\emailAdd{nicholas.j.sale@gmail.com}

\abstract{Topological Data Analysis (TDA) is a field that leverages tools and ideas from algebraic topology to provide robust methods for analysing geometric and topological aspects of data. One of the principal tools of TDA, persistent homology, produces a quantitative description of how the connectivity and structure of data changes when viewed over a sequence of scales. We propose that this presents a means to directly probe topological objects in gauge theories. We present recent work on using persistent homology to detect center vortices in SU(2) lattice gauge theory configurations in a gauge-invariant manner. We introduce the basics of persistence, describe our construction, and demonstrate that the result is sensitive to vortices. Moreover we discuss how, with simple machine learning, one can use the resulting persistence to quantitatively analyse the deconfinement transition via finite-size scaling, providing evidence on the role of vortices in relation to confinement in Yang-Mills theories.}

\FullConference{%
The 39th International Symposium on Lattice Field Theory,\\
8th-13th August, 2022,\\
Rheinische Friedrich-Wilhelms-Universität Bonn, Bonn, Germany
}

\usepackage{mathtools}
\newcommand{\reals}{\mathbb{R}}
\newcommand{\integers}{\mathbb{Z}}

\usepackage{dcolumn}
\newcolumntype{P}[1]{>{\centering\arraybackslash}p{#1}}

\newcommand{\twistB}{2.291}
\newcommand{\twistBE}{0.019}
\newcommand{\twistN}{0.614}
\newcommand{\twistNE}{0.079}

\newcommand{\linknnBfour}{2.2989}
\newcommand{\linknnBEfour}{0.0009}
\newcommand{\knnBfour}{2.2988}
\newcommand{\knnBEfour}{0.0007}
\newcommand{\knnNfour}{0.634}
\newcommand{\knnNEfour}{0.014}

\begin{document}
\maketitle

\section{Introduction}

Center vortices are topological defects that are observed in lattice quantum chromodynamics (QCD) simulations \cite{Biddle:2022zgw}. They provide a potential explanation for the mechanism of confinement in QCD and the deconfinement phase transition \cite{THOOFT19781, CORNWALL1979392, tHooft:1979rtg}, but existing methods to study them in configurations rely on gauge fixing and projection \cite{universe7050122} which suffers from the Gribov copies problem \cite{GRIBOV19781,STACK2001529}. In this project we leverage persistent homology \cite{Edelsbrunner2002TopologicalPA}, a tool from topological data analysis (TDA), to study center vortices in a gauge invariant manner. Rather than working with lattice QCD, we consider the $\mathrm{SU}(2)$ lattice gauge theory as a toy model since this also exhibits center vortices and a deconfinement phase transition.

\section{Lattice Model and Center Vortices}

The 4D $\mathrm{SU}(2)$ lattice gauge theory is specified by $\mathrm{SU}(2)$-valued variables $U_\mu(x)$, taking the form of a $2 \times 2$ complex matrix, located on each link $(x, \mu)$ of an $N_t \times N_s^3$ lattice $\Lambda$ with periodic boundary conditions, where $\mu \in \{ 0, 1, 2, 3 \}$ describes the direction in which the link emanates from the lattice site $x \in \Lambda$. Gauge invariant observables are obtained as traces of products of the link variables along closed paths $C$, also known as Wilson loops $W(C)$. The simplest non-trivial example is the Wilson loop around a $1 \times 1$ plaquette $(x, \mu, \nu)$ of the lattice:
$$W_{\mu, \nu}(x) = \frac{1}{2} \, tr \Big[ U_\mu(x) \, U_\nu(x + \hat{\mu}) \, U_\mu^\dagger(x + \hat{\nu}) \, U_\nu^\dagger(x) \Big].$$
We use this to define the Wilson action given a configuration $\mathbf{U} = \{ U_\mu(x) \}_{(x,\mu)}$ as
\begin{equation}
   S(\mathbf{U}) = - \frac{\beta}{4} \sum_{x, \mu < \nu} W_{\mu, \nu}(x) \label{eqn:action}
\end{equation}
where $\beta = 4/g^2$ and $g$ is the gauge coupling parameter. This in turn allows us to define the vacuum expectation value of any given observable $A(\mathbf{U})$ as
\begin{equation}
\langle A \rangle = \frac{\int d\mathbf{U} \, A(\mathbf{U}) \, e^{-S(\mathbf{U})}}{\int d\mathbf{U} \, e^{-S(\mathbf{U})}} \label{eqn:vev}
\end{equation}
where $d\mathbf{U} = \prod_{x,\mu} dU_\mu(x)$ is a product of Haar measures over $\mathrm{SU}(2)$ for each link variable. In practice we estimate expectations using Monte Carlo methods, where Eq. \eqref{eqn:vev} becomes a simple mean of the observed values.

We introduce center vortices following \cite{THOOFT19781}. Fix a time slice at time $t$. Given two closed oriented curves $C$ and $C^{\prime}$ in that 3-dimensional slice with linking number $m$, a loop operator $B(C^{\prime}, t)$ can be defined that has the following commutation algebra with the Wilson loop $W(C, t)$: 
\begin{equation}
    W({C}, t) B({C^{\prime}}, t) - (-1)^m B({C^{\prime}}, t) W({C}, t) = 0.
\end{equation}
The operator $B(C^{\prime}, t)$ is called the 't Hooft loop which, when acting on a gauge configuration, creates a magnetic flux with the resulting observable effect of multiplication of all Wilson loops around curves ${C}$ with linking number 1 with ${C^{\prime}}$ by $-1$. The 't Hooft loop is therefore said to be a vortex creation operator. Since the center of the group $Z(\mathrm{SU}(2)) = \{ I, -I \} \cong \integers_2$ plays a role (as exposed by the factor $(-1)^m$), the vortices are called center vortices. Allowing the curve $C^\prime$ to vary continuously over time slices, we see that a vortex traces out a surface in 4-space, closed by the periodic boundary conditions.

The above describes a 'thin' vortex. In practice center vortices have some finite thickness, so that only larger Wilson loops may fully link with them and obtain the full center charge. Loops that partially link may still obtain a partial charge, being multiplied by some matrix lying between $I$ and $-I$ in $\mathrm{SU}(2)$.

To explicitly insert a thin vortex into the system to study we will make use of the trick of imposing twisted boundary conditions \cite{THOOFT1979141}. The idea is that we negate the contribution to the action of the co-closed collection of plaquettes
$$T = \{ ((0,0,y,z), 0, 1) \,\, \vert \,\, 0 \leq y, z < N_s \}$$
corresponding to a surface wrapping round the latter two spatial dimensions of the lattice. The action with twisted boundary conditions becomes
\begin{equation}
    S_T(\mathbf{U}) = - \frac{\beta}{4} \Bigg[\sum_{\substack{x,\mu < \nu \\ (x, \mu, \nu) \not\in T}} W_{x, \mu, \nu} - \sum_{\substack{x,\mu < \nu \\ (x, \mu, \nu) \in T}} W_{x, \mu, \nu}\Bigg]
    \label{eqn:tbc_action}
\end{equation}
which we refer to as the twisted action. This modification of the action allows the lattice to support an odd number of center vortices wrapping in the $yz$ plane, which is not allowed by the usual periodic boundary conditions of the Wilson action. We can think of this twisted action as explicitly inserting a thin vortex into the system on the surface defined by $T$, so that the system is forced to generate a (thick) vortex to cancel it out. We shall denote expectations calculated with respect to this twisted action by $\langle A \rangle_{\text{twist}}$, where $A$ is a generic observable.

\section{Persistent Homology}

Persistent homology takes a nested sequence of topological spaces and produces a topological summary called a barcode or persistence diagram \cite{Edelsbrunner2002TopologicalPA} (see \cite{carlsson2020persistent, ph_survey_edels_harer, otter, ghrist_barcodes} for useful references). In this case we assign a filtered cubical complex $F_{\boldsymbol{U}} : \reals \rightarrow \text{CubicalComplex}$ (such that $s \leq t$ implies $F(s) \subseteq F(t)$) to each configuration $\boldsymbol{U} = \{U_\sigma(x)\}$ and compute the persistent homology of this filtered cubical complex. Our construction is based on Wilson loops and therefore yields gauge-invariant persistence diagrams.

The idea is to explicitly construct a cubical model of vortex surfaces, under the assumption that vortices are thin. Vortex sheets live on the dual lattice $\Lambda^*$. We therefore consider a decomposition of the spacetime manifold into a cubical complex $Y$ given by $\Lambda^*$ in which there is a vertex for each dual lattice site, an edge for each link in the dual lattice, a 2-cube for each dual plaquette, etc. Note that there is a bijection between the 2-cubes of this cubical complex (i.e. dual plaquettes) with the plaquettes of the original lattice $\Lambda$, pairing a plaquette with the dual plaquette that intersects it at a single point. We will construct a filtration of $Y$ by letting each 2-cube enter at a filtration index given by the value of the Wilson loop around the plaquette it is paired with in the bijection.

To define the filtered complex we give a filtration index $f(c) \in \reals$ for each cube $c$ in $Y$ specifying when it appears. Then $F_{\boldsymbol{U}}(s)$ is the subcomplex of $Y$ consisting of all cubes $c$ for which $f(c) \leq s$. That is, $$F_{\boldsymbol{U}}(s) = f^{-1}(-\infty, s].$$ Since we are attempting to model vortex surfaces, we will initially specify when the 2-cubes are to enter the filtered complex and then introduce the cubes of other dimensions based on these.

Our construction of the function $f$ is the following:
\begin{enumerate}
    \item For each 2-cube $c$, i.e. dual plaquette, we set $f(c)$ equal to the value of the Wilson loop around the plaquette paired with it by the bijection.
    \item Since a 2-cube is not allowed to be included before its constituent 1-cubes and 0-cubes in a cubical complex, setting $f(c)$ for these to be the smallest value of $f$ of any of the 2-cubes they are incident to.
    \item For the 3-cubes and 4-cubes we follow a clique-like rule, setting $f(c)$ for these to be the largest value of $f$ of any of the 2-cubes contained in their boundary.
\end{enumerate}
Thus for $s < -1$, $F_{\boldsymbol{U}}(s)$ is the empty complex and for $s \geq 1$, $F_{\boldsymbol{U}}(s)$ is the filled in tiling of spacetime, homeomorphic to a 4-torus due to the periodic boundary conditions. Going between these values, the first cubes to enter $F_{\boldsymbol{U}}$ are surfaces made up of plaquettes in bijection with Wilson loops that are close to $-1$. The idea therefore is that thin vortex surfaces will enter the filtered complex early. Moreover, since small Wilson loops like those considered here still pick up a partial charge from thick vortices, surfaces representing those thick vortices ought to enter the filtered complex earlier than they otherwise would have. We expect to detect these closed surfaces in persistent $H_2$. We may also see other topological features such as the presence of handles or holes in $H_1$, as well as the transient low-persistence points in persistent $H_0$ and $H_1$ that arise as the surface forms near the start of the filtration.

\section{Detecting Vortices with Twisted Boundary Conditions}

We first test if the persistent homology can distinguish between configurations generated using the Wilson action and configurations generated using the twisted action. That is, if it detects an inserted vortex. 

For $N_s \in \{ 12, 16, 20 \}$, fixing $N_t = 4$, we generate $200$ configurations using the Wilson action (\ref{eqn:action}) and $200$ configurations using the twisted action (\ref{eqn:tbc_action}) for each $\beta \in \{ 1.5, 1.6, \dots 2.9 \}$. Configurations are generated using the HiRep software \cite{PhysRevD.81.094503} with 1 heatbath step and 4 overelaxation steps for each Monte Carlo step and a sample taken every 100 Monte Carlo steps.

\begin{figure}[h]
    \centering
    \scalebox{0.5}{\includegraphics{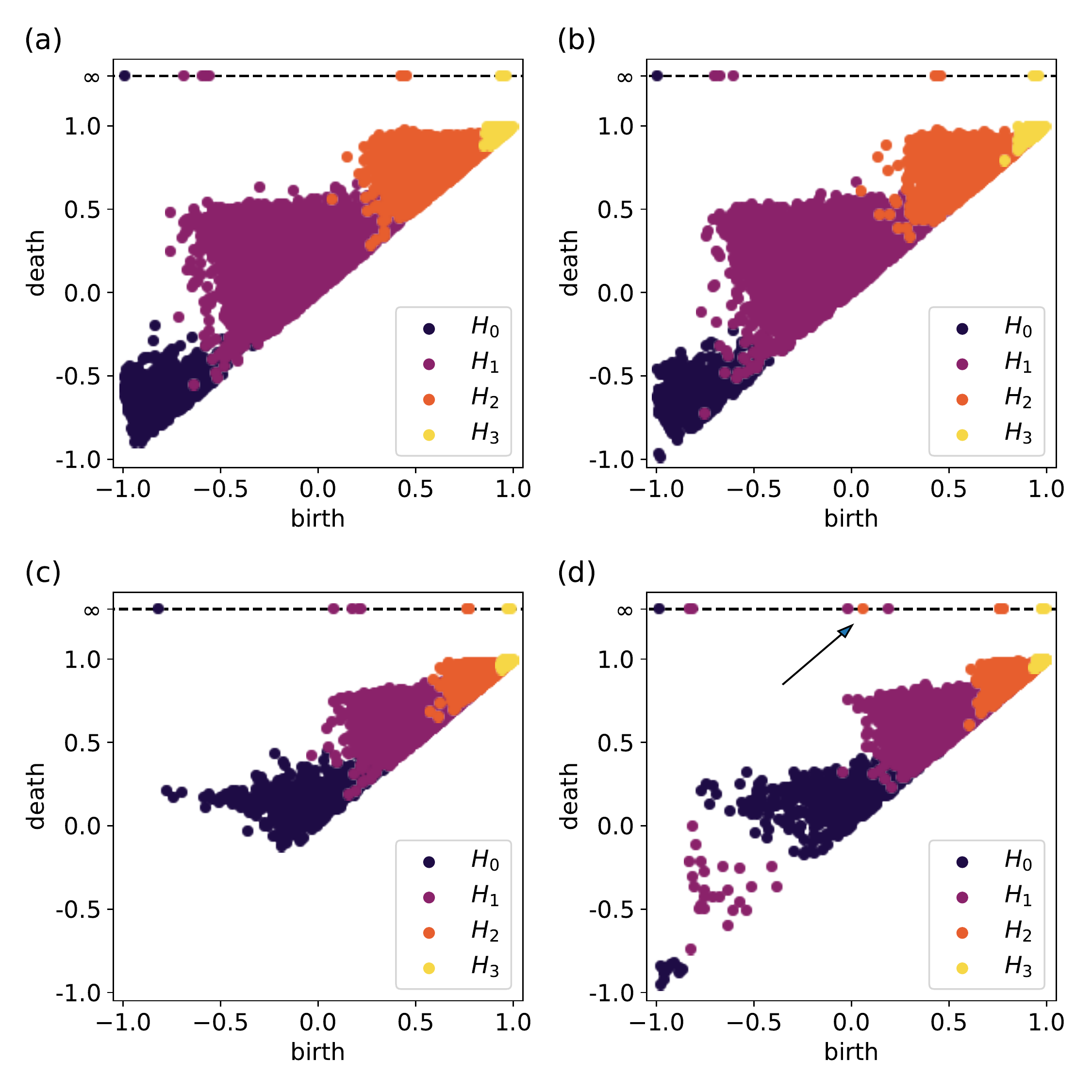}}
    \caption{Sample persistence diagrams of individual configurations obtained using the following actions and values of $\beta$: (a) Wilson, $\beta = 1.5$ (b) twisted, $\beta = 1.5$ (c) Wilson, $\beta = 2.9$ (d) twisted, $\beta = 2.9$. The arrow in (d) indicates the point $(b, \infty) \in PH_2$ with the smallest birth index $b$. Note the distance between it and the others.}
    \label{fig:PDs}
\end{figure}

Figure \ref{fig:PDs} shows example persistence diagrams obtained using the two different actions and in the two phases of the model.

\begin{figure}
\centering
\begin{subfigure}{.5\textwidth}
  \centering
  \scalebox{0.5}{\includegraphics{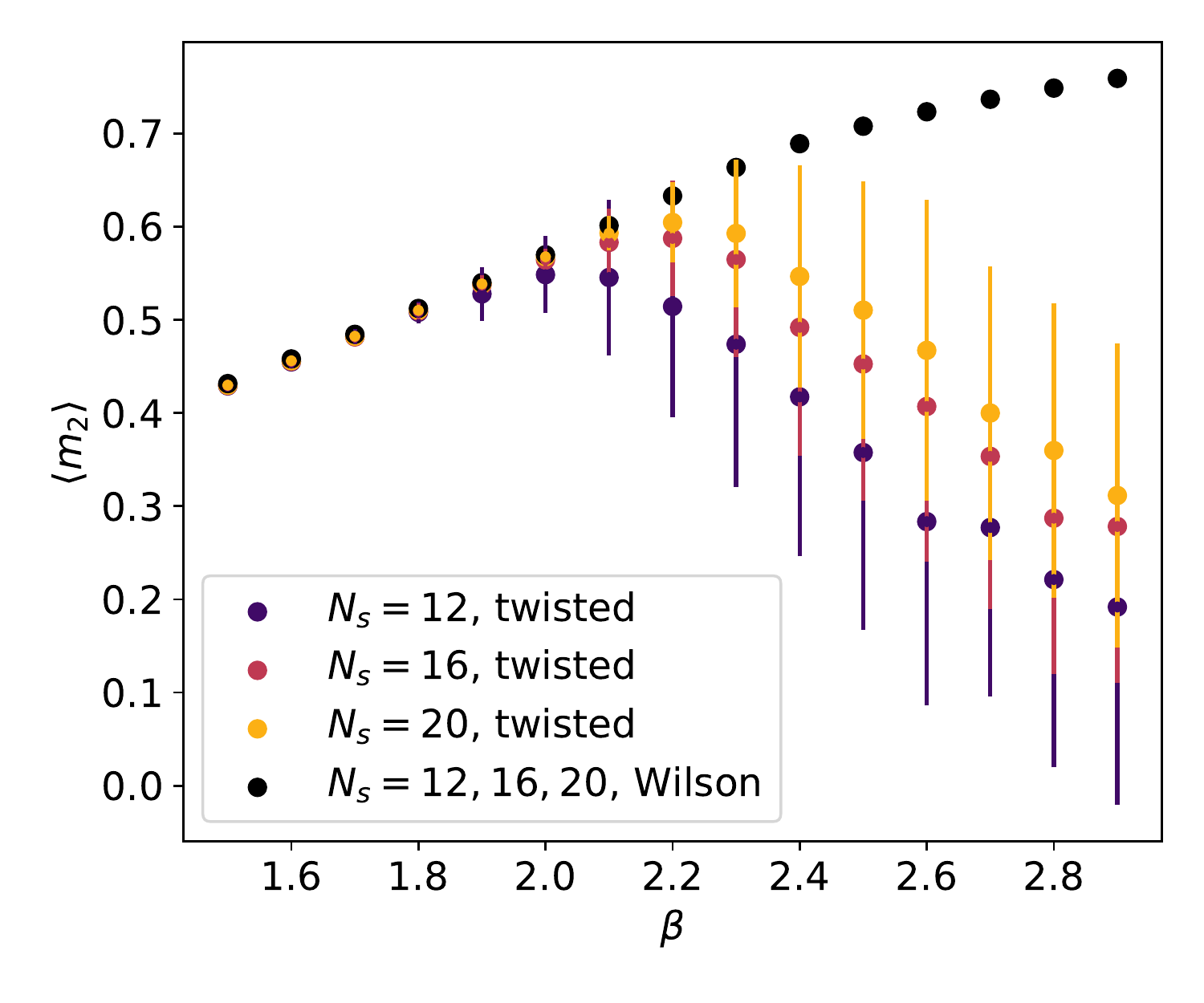}} 
  \caption{}
  \label{fig:m2}
\end{subfigure}%
\begin{subfigure}{.5\textwidth}
  \centering
  \scalebox{0.5}{\includegraphics{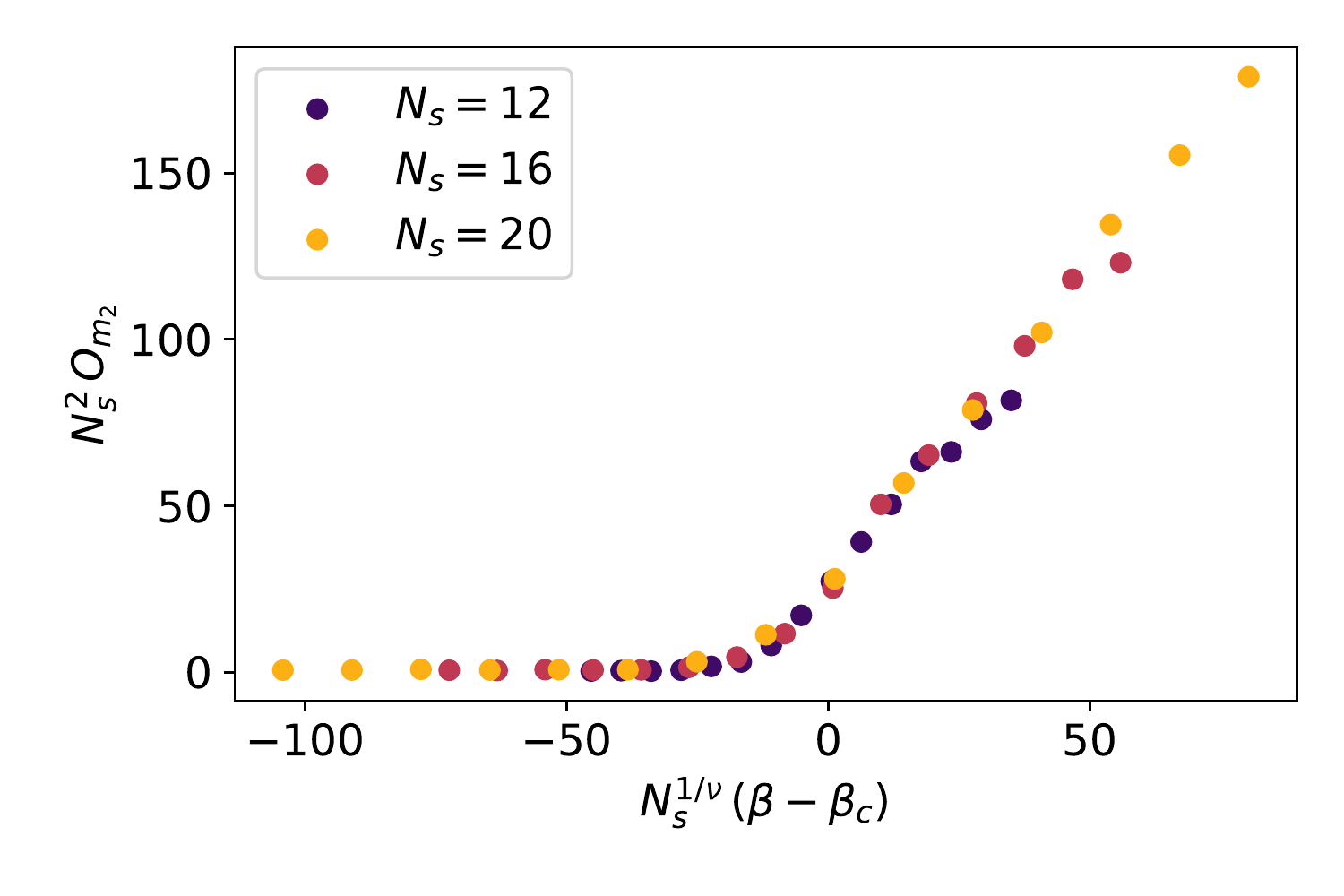}}
  \caption{}
  \label{fig:O_m2}
\end{subfigure}
\caption{(a) The expected value of the observable $m_2$ as a function of $\beta$ plotted for different values of $N_s$ and with the Wilson and twisted actions. (b) The curve collapse of $O_{m_2}$ using $\beta_c = \twistB$ and $\nu = \twistN$. Error bars are not shown for clarity but are comparable to those in (a).}
\end{figure}

In the deconfined phase, one of the infinite death points in $H_2$ is born much earlier for the twisted action. This represents a surface which wraps the periodic boundary conditions of the lattice entering our filtration early: i.e. the inserted vortex. We therefore define the following observable based on the persistence diagram of a configuration
$$m_2 = \min \big\{ \, b \,\, \big\vert \,\, (b,\infty) \in PH_2 \, \big\}.$$
The expected value of $m_2$ for different lattice sizes with the Wilson action and twisted action are shown in Figure \ref{fig:m2}. Note that there is no difference between the expectations estimated using the different actions well into the confined phase, but in the deconfined phase the curves separate. As the lattice size increases, the point at which the curves diverge approaches the critical $\beta$ of the phase transition from below. These observations motivate measuring the difference between the expected values using different actions
$$O_{m_2} = \langle m_2 \rangle - \langle m_2 \rangle_{\text{twist}}$$
as a phase indicator which will be zero in the confined phase and non-zero in the deconfined phase, similar to the definition of an order parameter but without the requirement to detect any symmetry breaking. A finite-size scaling analysis of this quantity yields the curve collapse in Figure \ref{fig:O_m2}, computed numerically using the Nelder-Mead method following \cite{Bhattacharjee2001AMO}. The resulting estimates of $\beta_c$ and $\nu$,
\begin{align*}
    \beta_c &= \twistB \pm \twistBE\\
    \nu &= \twistN \pm \twistNE,
\end{align*}
 are in agreement with our reference estimate $\beta_c = 2.2986(6)$ from \cite{lucini_SU(N)_transition} and $\nu = 0.629971(4)$ from \cite{precision_ising}. Error estimates are obtained by performing $2000$ bootstraps.

\section{Detecting the Deconfinement Transition Without Twisted Boundary Conditions}

Using a machine learning framework inspired by that in \cite{sale2022quant}, we investigate if it is possible to extract the critical $\beta$ and critical exponent $\nu$ of the deconfinement transition for $N_t = 4$ using configurations sampled using the Wilson action alone.

\begin{figure}[h]
    \centering
    \scalebox{0.7}{\includegraphics{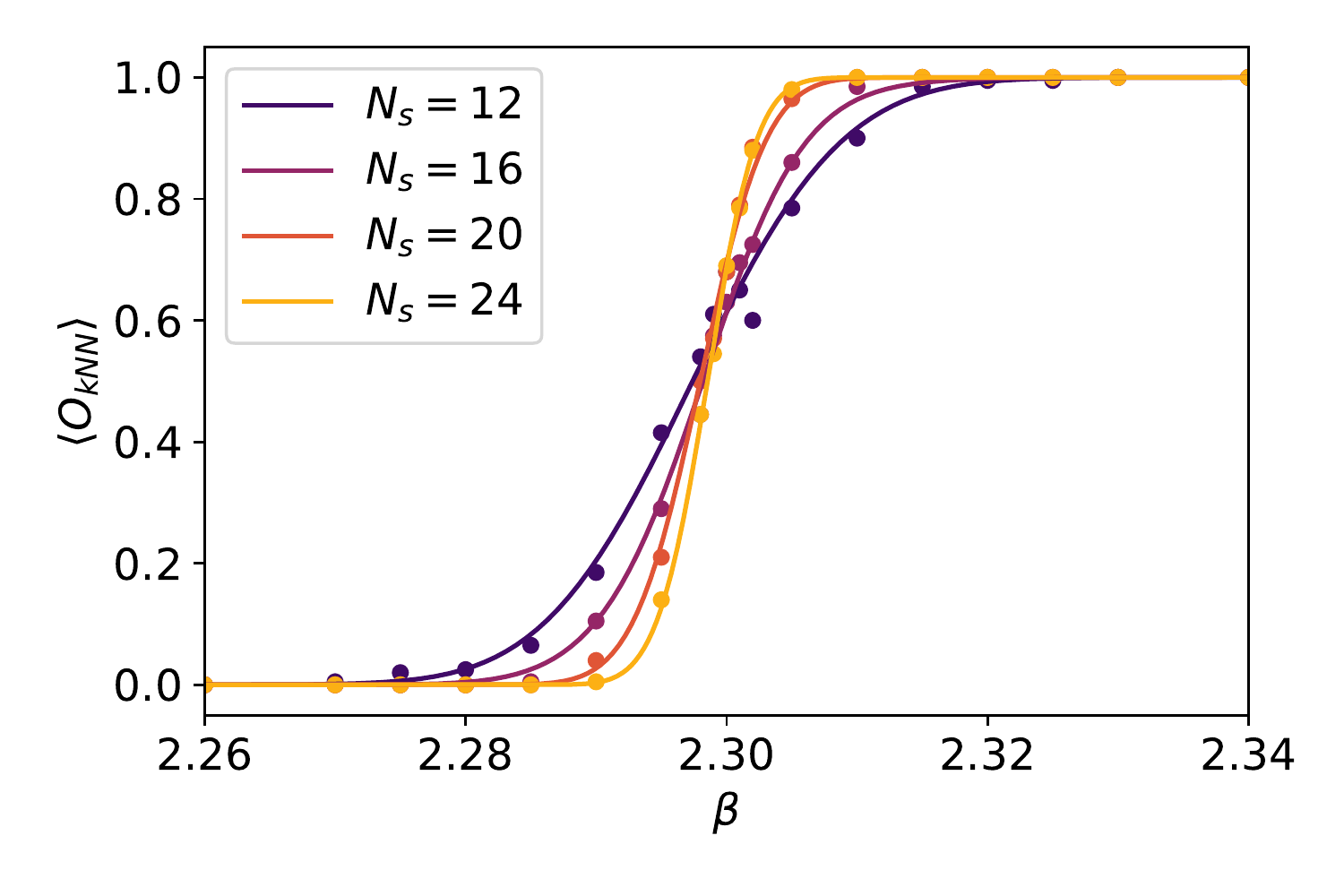}}
    \caption{Plot showing our phase indicator $\langle O_{k\mathrm{NN}} \rangle$ as a function of $\beta$ for $N_t = 4$. The points show the measured expectations and the curve is the output of histogram reweighting these measurements.}
    \label{fig:Oknn}
\end{figure}

For lattices of size $4 \times N_s^3$ with $N_s \in \{ 12, 16, 20, 24 \}$, we train a $k$-nearest neighbours classifier ($k = 30$) on the concatenated $PH_0$, $PH_1$, $PH_2$ and $PH_3$ persistence images of $200$ configurations sampled at each $\beta$ in the confined and deconfined regions given in Table \ref{tab:Nt4_betas}. The classifier is then used to produce a predicted classification $O_{k\mathrm{NN}}$ for $200$ configurations sampled for each value of $\beta$ in the critical region. The resulting curve is shown in Figure \ref{fig:Oknn}.

\begin{table}[h!]
\centering
\begin{tabular}{ || P{5em} | P{19em} || }
 \hline
 Region & $\beta$ \\
 \hline
 Confined & 2.2 , 2.21, 2.22, 2.23, 2.24 \\ \hline

 Deconfined & 2.36, 2.37, 2.38, 2.39, 2.4 \\ \hline

 Critical & 2.25, 2.26, 2.27, 2.275, 2.28, 2.285, 2.29, 2.295, 2.298, 2.299, 2.3, 2.301, 2.302, 2.305, 2.31, 2.315, 2.32, 2.325, 2.33, 2.34, 2.35 \\
 \hline
\end{tabular}
\caption{Values of $\beta$ sampled at for the $N_t = 4$ phase transition.}
\label{tab:Nt4_betas}
\end{table}

\begin{figure}
\centering
\begin{subfigure}{.5\textwidth}
  \centering
  \scalebox{0.5}{\includegraphics{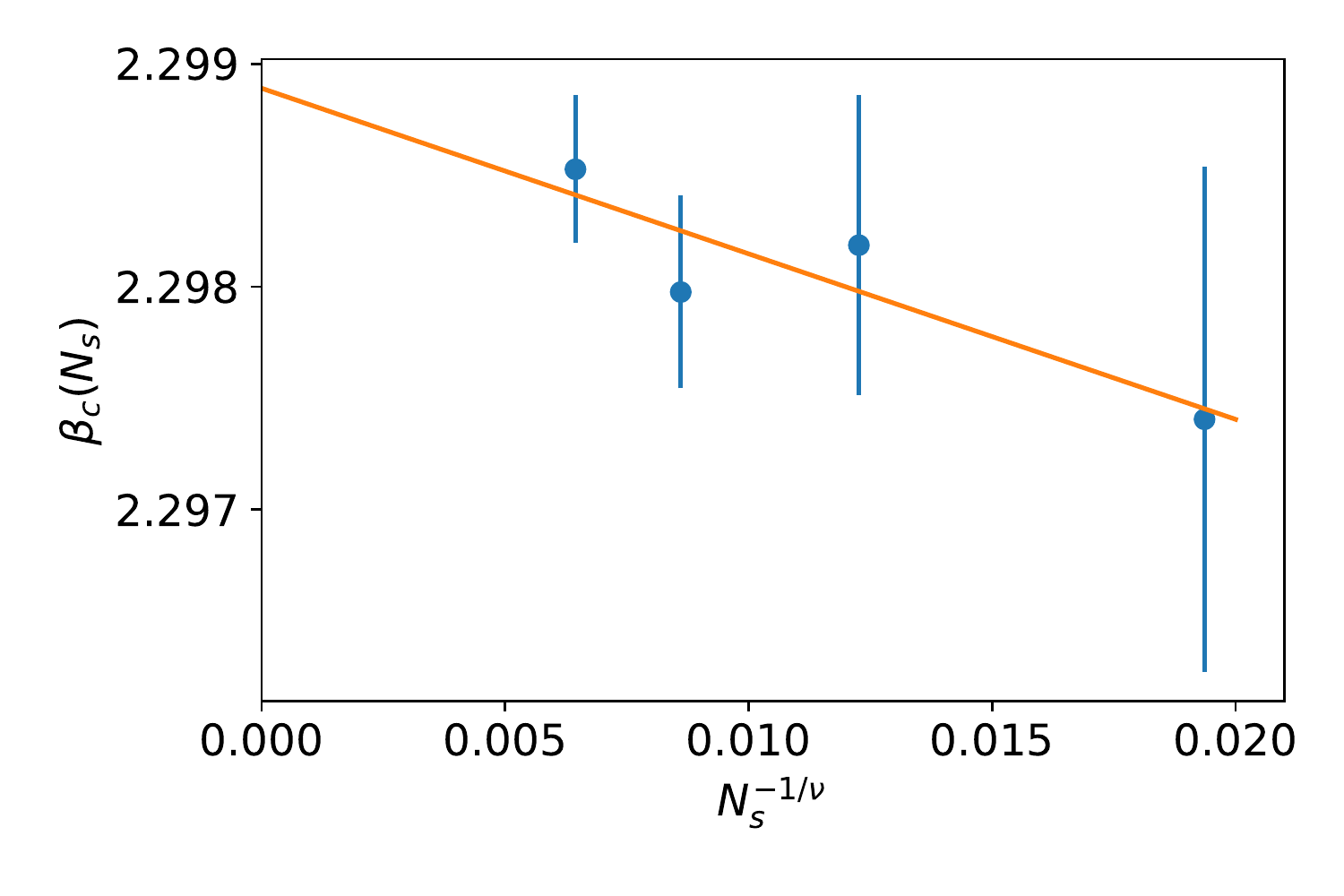}}
  \caption{}
  \label{fig:bc_line_Nt4}
\end{subfigure}%
\begin{subfigure}{.5\textwidth}
  \centering
  \scalebox{0.5}{\includegraphics{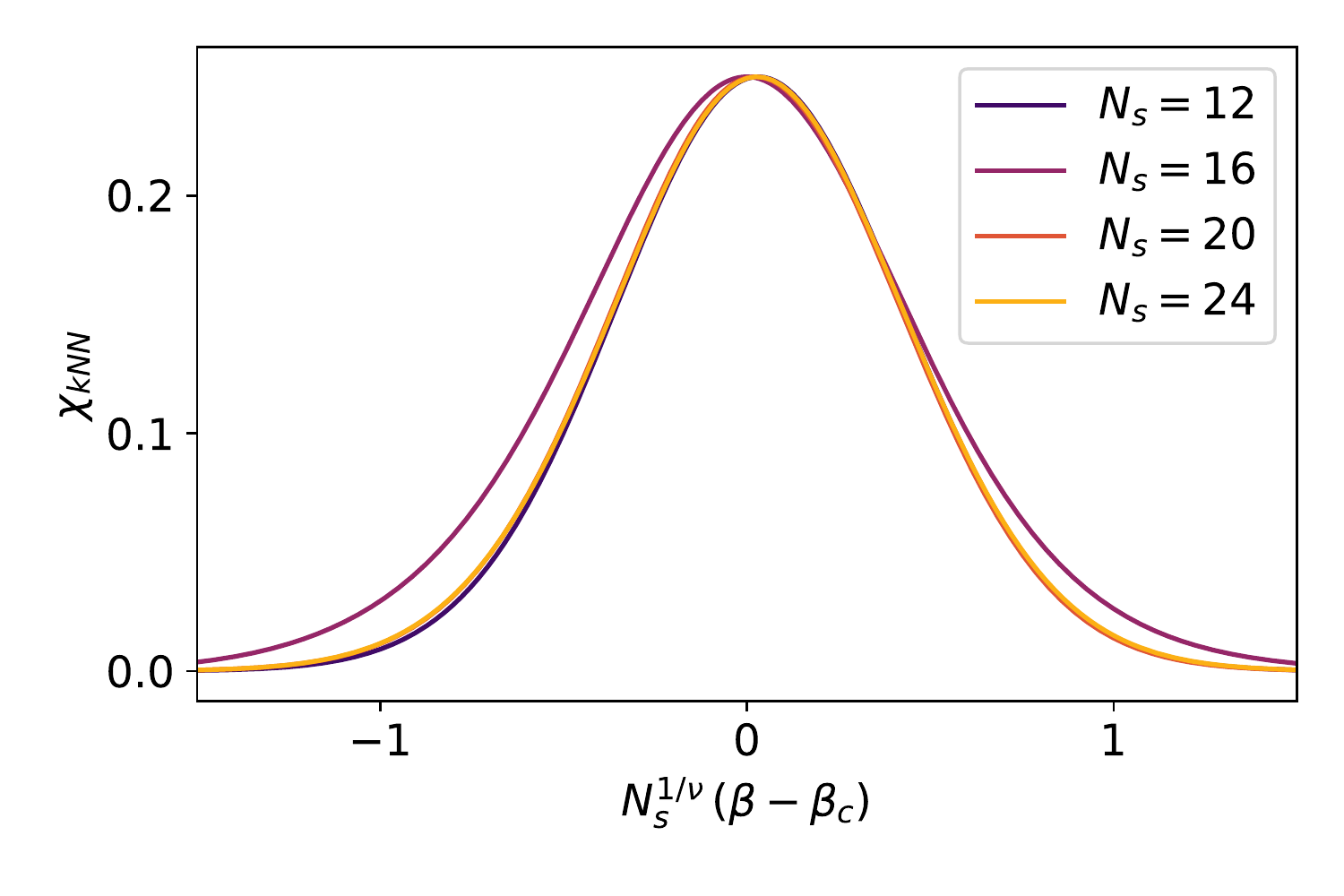}}
  \caption{}
  \label{fig:collapsed_Xknn_Nt4}
\end{subfigure}
\caption{(a) Estimating $\beta_c$ using a linear fit, assuming known $\nu$. The pseudo-critical values of $\beta$, obtained as the points where the curves in Figure \ref{fig:Oknn} cross $0.5$, are fitted to a straight line against $N_s^{-1/\nu}$ with $\nu = 0.629971$. Error bars are estimated by bootstrapping. (b) The curve collapse of $\chi_{k\mathrm{NN}}$ using $\beta_c = \knnBfour$ and $\nu = \knnNfour$.}
\label{fig:test}
\end{figure}

Assuming a known value of $\nu = 0.629971$, we can do a finite-size scaling analysis by extracting the pseudo-critical point for each $N_s$ via the implicit formula $\langle O_{k\mathrm{NN}} \rangle (\beta_c(N_s)) = 0.5$ then fitting these to the straight line ansatz $\beta_c(N_s) - \beta_c(\infty) \propto N_s^{-1/\nu}$ to extract $\beta_c = \beta_c(\infty)$. The resulting fit is shown in Figure \ref{fig:bc_line_Nt4}. The intercept yields $\beta_c = \linknnBfour \pm \linknnBEfour$, supporting our reference estimate of $\beta_c = 2.2986(6)$ from \cite{lucini_SU(N)_transition}.

Alternatively, we can try to estimate both $\beta_c$ and $\nu$ simultaneously via a curve collapse of the variance curves $\chi_{k\mathrm{NN}} = \langle O_{k\mathrm{NN}}^2 \rangle - \langle O_{k\mathrm{NN}} \rangle^2$ using a numerical procedure like that in \cite{Bhattacharjee2001AMO}. The result of using the Nelder-Mead method is shown in Figure \ref{fig:collapsed_Xknn_Nt4}. The obtained estimates of $\beta_c$ and $\nu$
\begin{align*}
    \beta_c &= \knnBfour \pm \knnBEfour\\
    \nu &= \knnNfour \pm \knnNEfour
\end{align*}
are consistent with previous estimates.

\section{Conclusion}

We designed a methodology to use persistent homology to detect center vortices and tested the efficacy of this by using it to distinguish configurations generated using a twisted action from configurations generated using the usual Wilson action. We also performed a quantitative analysis of the deconfinement transition using two different phase indicators derived from the persistent homology. We argue that, since the methodology summarises center vortices but is also sensitive to the phase transition, then center vortices must play some role in the phase transition. For a stronger argument we would need to consider the sensitivity of our methodology to the objects involved in other pictures of confinement, e.g., monopoles.

\begin{acknowledgments}
Numerical simulations have been performed on the Swansea SUNBIRD system. This system is part of the Supercomputing Wales project, which is part-funded by the European Regional Development Fund (ERDF) via Welsh Government. Configurations of the $\mathrm{SU}(2)$ lattice gauge theory were sampled using the HiRep software \cite{PhysRevD.81.094503}. Persistent homology calculations were performed using giotto-tda \cite{tauzin2020giottotda}. Histogram reweighting calculations were performed using pymbar \cite{Shirts2008StatisticallyOA}. NS has been supported by a Swansea University Research Excellence Scholarship (SURES). JG was supported by EPSRC grant EP/R018472/1. BL received funding from the European Research Council (ERC) under the European Union’s Horizon 2020 research and innovation programme under grant agreement No 813942. The work of BL was further supported in part by the UKRI Science and Technology Facilities Council (STFC) Consolidated Grant ST/T000813/1, by the Royal Society Wolfson Research Merit Award WM170010 and by the Leverhulme Foundation Research Fellowship RF-2020-461{\textbackslash}9. 
\end{acknowledgments}

\bibliographystyle{hplain}
\bibliography{bib}

\begin{thebibliography}{10}

\bibitem{Bhattacharjee2001AMO}
S.~M. Bhattacharjee and F.~Seno.
\newblock A measure of data collapse for scaling.
\newblock {\em Journal of Physics A}, 34:6375--6380, 2001.

\bibitem{Biddle:2022zgw}
James Biddle, Waseem Kamleh, and Derek Leinweber.
\newblock {Static quark potential from centre vortices in the presence of
  dynamical fermions}, 2022, arXiv:2206.00844.

\bibitem{carlsson2020persistent}
Gunnar Carlsson.
\newblock Persistent homology and applied homotopy theory, 2020,
  arXiv:2004.00738.

\bibitem{CORNWALL1979392}
John~M. Cornwall.
\newblock Quark confinement and vortices in massive gauge-invariant qcd.
\newblock {\em Nuclear Physics B}, 157(3):392--412, 1979.

\bibitem{PhysRevD.81.094503}
Luigi Del~Debbio, Agostino Patella, and Claudio Pica.
\newblock Higher representations on the lattice: Numerical simulations, su(2)
  with adjoint fermions.
\newblock {\em Phys. Rev. D}, 81:094503, 2010.

\bibitem{Edelsbrunner2002TopologicalPA}
H.~Edelsbrunner, D.~Letscher, and A.~Zomorodian.
\newblock Topological persistence and simplification.
\newblock {\em Discrete \& Computational Geometry}, 28:511--533, 2002.

\bibitem{ph_survey_edels_harer}
Herbert Edelsbrunner and John Harer.
\newblock Persistent homology—a survey.
\newblock {\em Discrete \& Computational Geometry - DCG}, 453, 01 2008.

\bibitem{ghrist_barcodes}
Robert Ghrist.
\newblock Barcodes: The persistent topology of data.
\newblock {\em BULLETIN (New Series) OF THE AMERICAN MATHEMATICAL SOCIETY}, 45,
  02 2008.

\bibitem{universe7050122}
Rudolf Golubich and Manfried Faber.
\newblock A possible resolution to troubles of su(2) center vortex detection in
  smooth lattice configurations.
\newblock {\em Universe}, 7(5), 2021.

\bibitem{GRIBOV19781}
V.N. Gribov.
\newblock Quantization of non-abelian gauge theories.
\newblock {\em Nuclear Physics B}, 139(1):1--19, 1978.

\bibitem{precision_ising}
Filip Kos, David Poland, David Simmons-Duffin, and Alessandro Vichi.
\newblock Precision islands in the ising and $o(n)$ models.
\newblock {\em Journal of High Energy Physics}, 2016.

\bibitem{lucini_SU(N)_transition}
Biagio Lucini, Michael Teper, and Urs Wenger.
\newblock The high temperature phase transition in su(n) gauge theories.
\newblock {\em Journal of High Energy Physics}, 2004, 07 2003.

\bibitem{otter}
Nina Otter, Mason Porter, Ulrike Tillmann, Peter Grindrod, and Heather
  Harrington.
\newblock A roadmap for the computation of persistent homology.
\newblock {\em EPJ Data Science}, 6, 2015.

\bibitem{sale2022quant}
Nicholas Sale, Jeffrey Giansiracusa, and Biagio Lucini.
\newblock Quantitative analysis of phase transitions in two-dimensional $xy$
  models using persistent homology.
\newblock {\em Phys. Rev. E}, 105:024121, 2022.

\bibitem{Shirts2008StatisticallyOA}
Michael~R. Shirts and John~D. Chodera.
\newblock Statistically optimal analysis of samples from multiple equilibrium
  states.
\newblock {\em The Journal of chemical physics}, 129 12:124105, 2008.

\bibitem{STACK2001529}
John~D. Stack and William~W. Tucker.
\newblock The gribov ambiguity for maximal abelian and center gauges in su(2)
  lattice gauge theory.
\newblock {\em Nuclear Physics B - Proceedings Supplements}, 94(1):529--531,
  2001.
\newblock Proceedings of the XVIIIth International Symposium on Lattice Field
  Theory.

\bibitem{THOOFT19781}
G.~{'t Hooft}.
\newblock On the phase transition towards permanent quark confinement.
\newblock {\em Nuclear Physics B}, 138(1):1--25, 1978.

\bibitem{THOOFT1979141}
G.~{'t Hooft}.
\newblock A property of electric and magnetic flux in non-abelian gauge
  theories.
\newblock {\em Nuclear Physics B}, 153:141--160, 1979.

\bibitem{tHooft:1979rtg}
Gerard 't~Hooft.
\newblock {A Property of Electric and Magnetic Flux in Nonabelian Gauge
  Theories}.
\newblock {\em Nucl. Phys. B}, 153:141--160, 1979.

\bibitem{tauzin2020giottotda}
Guillaume Tauzin, Umberto Lupo, Lewis Tunstall, Julian~Burella Pérez, Matteo
  Caorsi, Anibal Medina-Mardones, Alberto Dassatti, and Kathryn Hess.
\newblock giotto-tda: A topological data analysis toolkit for machine learning
  and data exploration, 2020, arXiv:2004.02551.

\end{thebibliography}

\end{document}